\documentclass{article}

\usepackage{arxiv}
\usepackage{timet}
\usepackage{amsmath, enumitem}
\usepackage{amsfonts}
\usepackage{upgreek}
\usepackage{graphicx}
\graphicspath{{../Figures/}} 
\usepackage{setspace}
\usepackage{multirow}
\usepackage[T1]{fontenc}
\usepackage{xcolor}
\usepackage[skins,breakable]{tcolorbox}

\usepackage{timet}

\usepackage[utf8]{inputenc} 
\usepackage[T1]{fontenc}    
\usepackage{hyperref}       
\usepackage{url}            
\usepackage{booktabs}       
\usepackage{amsfonts}       
\usepackage{nicefrac}       
\usepackage{microtype}      
\usepackage{graphicx}
\usepackage{subcaption}
\usepackage{lipsum}
\usepackage{natbib}
\usepackage{mathrsfs}
\usepackage{amsmath}
\usepackage{bm}
\usepackage{gensymb}

\usepackage{dirtytalk}

\DeclareMathOperator{\NormalDist}{\mathcal{N}}

\newcommand{\bs}{\boldsymbol}

\newtcolorbox{cvbox}[2][]{%
  blanker,
  width = 0.9\textwidth,
  after skip=8mm,
  title=#2,
  breakable,
  #1
}

\title{Introducing Conceptual Geological Information into Bayesian Tomographic Imaging}
\author{Hugo Bloem \\
        School of Geosciences \\
	University of Edinburgh\\
	Edinburgh, Unite Kingdom \\
	\texttt{hugo.bloem@ed.ac.uk} \\
	\And
	Andrew Curtis \\
	School of Geosciences \\
	University of Edinburgh\\
	Edinburgh, Unite Kingdom \\
	\texttt{andrew.curtis@ed.ac.uk} \\
	\And
	Daniel Tetzlaff \\
	Westchase Software Corporation \\
	\texttt{dan@westchasesoftware.com} \\}

\begin{document}
	\maketitle
        

		 
        

        


	

	\begin{abstract}Geological process models typically simulate a range of dynamic processes to evolve a base topography into a final 2-dimensional cross-section or 3-dimensional geological scenario. In principle, process parameters may be updated to better align with observed geophysical or geological data; however, many realisations of process models that embody different conceptual models may provide similar consistency with observed data, and finding all such realisations may be infeasible due to the computational demands of the task. Alternatively, geophysical probabilistic tomographic methods may be used to estimate the family of models of a target subsurface structure that are consistent both with information obtained from previous experiments and with new data (the Bayesian posterior probability distribution). However, this family seldom embodies geologically reasonable images. We show that the posterior distribution of tomographic images obtained from travel time data can be enhanced by injecting geological prior information into Bayesian inference, and that we can do this near-instantaneously by trained Mixture Density Networks (MDNs). We invoke two geological concepts as prior information about the depositional environment of an imaged target structure: a braided river system, and a set of marine parasequences, each parameterised by a Generative Adversarial Network. Data from a target structure can then be used to infer the image parameter values using either geological concept using MDNs. Our MDN solutions closely resemble those found using expensive Monte Carlo methods, and while the use of incorrect geological conceptual models provides less accurate results the mean structures still approximate the target. We then show that a classifier neural network can infer the correct geological conceptual model. Thus, imposing even incorrect geological prior information may improve geophysical tomographic images compared to those obtained without geological prior information, and in principle geological conceptual models can be inferred directly from geophysical travel time data.
 	\end{abstract}

\section{Introduction} \label{chap::intro}
Geological process models simulate dynamic processes to transform an initial topography into a geologically plausible 2-dimensional cross section or 3-dimensional geological model \citep{tetzlaff1989simulating, paola2000quantitative, burgess2001numerical, hill2009modeling}. The input parameters to the geological process model may be updated to align the model output with geophysical data such as recorded seismic travel times or waveforms. However, a number of problems may occur: some geological process models are chaotic in their behaviour \citep{tetzlaff1989simulating, burgess2004sensitive} such that a small change in the input could lead to a significantly different output. In addition, the model output is never uniquely constrained by data so that an infinite family of dynamic models are consistent with observations. Finding the family of realisations that are consistent with observed data may be infeasible due to the computational expense involved. And finally, the true structure of the Earth always deviates from the model output, even when compared at the level of detail of the model. As an alternative, we may look to probabilistic inversion methods to identify the family of models that fit geophysical data to within their uncertainties \citep{tarantola2005inverse}. Unfortunately, as applied to-date, geophysical inversion methods do not impose geological realism as a criterion on the solution. As a result, the inferred models are usually geologically implausible. In this paper we aim to combine the geological prior information embodied within process models with probabilistic inversion methods of geophysics to obtain a set of geological models that both fit the recorded data and are geologically reasonable.

Geological process models employ numerical simulation of various dynamic processes to transform an initial topography into a geologically plausible 2-dimensional cross section or 3-dimensional geological model \citep{tetzlaff1989simulating}. The input parameters to the geological process model could be updated to align the output model with a geophysical data like seismic travel times. However, geological process models are chaotic in their behaviour \citep{tetzlaff1989simulating}, a small change in the input could lead to a significantly different model. Furthermore, finding multiple realisations that fit the data may be infeasible due to the computational expense. We therefore look to probabilistic inversion methods to find all models that fit the geophysical data according to some misfit value \citep{tarantola1982}. However, as the only criterion is the data misfit the found models may be geologically implausible. We aim to combine the geological plausibility of process models with probabilistic inversion to obtain a set of geological models that both fit the recorded data and are geological.

Seismic travel time tomography is commonly applied to image the Earth’s subsurface \citep{aki1976determination, dziewonski1987global, lee1995time, zhang2020seismic, tsekhmistrenko2021tree}. By measuring the time taken for waves to travel between pairs of points on the Earth's surface, tomographic methods estimate maps of subsurface velocities in up to three dimensions. The subsurface is usually described by a finite-dimensional parameter matrix $\bs{m}$ which often consists of seismic velocities and/or densities at each of a set of subsurface locations. We study the situation where those parameters are to be inferred from a vector of recorded data $\bs{d}$ which describe the travel times of the seismic energy between a set of sources and receivers. 

Estimating subsurface velocities from travel times is a nonlinear inverse problem \citep{aki1976determination}. The unknown inverse function is potentially complicated and ill-posed, and always has a non-unique solution which means that infinitely many subsurface parameter matrices fit the data to within their uncertainties. It is therefore impossible to infer which particular parameter matrix produced the recorded data; the most that one can achieve is to constrain the family of parameter matrices that are consistent with measured data as tightly as possible. 

Bayesian inversion works in this context by providing a general method to define the statistical distribution of parameter matrices that fit the data, and to assign the probability density of each parameter matrix given the data $\rho(\bs{m}|\bs{d})$, known as the posterior probability distribution (pdf) which here is referred to simply as the \textit{posterior}. Bayes theorem allows us to calculate the posterior as follows:
\begin{equation}
    \rho(\bs{m}|\bs{d})=\frac{\rho(\bs{d}|\bs{m})\rho(\bs{m})}{\rho(\bs{d})}
    \label{eq:bayes_theorem}
\end{equation}
where $\rho(\bs{d}|\bs{m})$ is called the \textit{likelihood} which is the probability of observing data $\bs{d}$ if parameter matrix $\bs{m}$ is true, $\rho(\bs{m})$ is the prior distribution of $\bs{m}$ (referred to as the \textit{prior}), and $\rho(\bs{d})$ is the marginal probability of the data post-experiment -- also called the \textit{evidence}. \cite{tarantola2005inverse} gives a clear exposition of how each of these terms should be defined. 

Our aim is to improve knowledge about the parameters. Following Bayes theorem this can be achieved by increasing the amount of relevant information in the observed data set which is represented by the likelihood, or increasing information in the prior pdf. The likelihood and prior have equal mathematical weight in Equation \ref{eq:bayes_theorem}, and while most work focuses on adding information through better data or improved data processing and thus targets the likelihood, in this study we aim to inject more geological information about the parameters through the prior probability distribution. This paper focuses on the development and demonstration of a methodology that introduces the information and solves the resulting inverse problem efficiently.

Conceptual models are the hierarchical highest level of information in most geological studies. In the current context they describe our understanding of which geological processes have influenced the current subsurface structure and composition. We consider two geological conceptual models in this study: sedimentary structures created either by terrestrial river channel systems, or alternatively by marine parasequences. We represent each conceptual model by large sets of different geometries of rock types that might be generated by processes invoked in that model. Each set of geometries is in turn represented by a group of neural networks which are trained to regress through each set, to allow other representative geometries to be generated efficiently. 

The generation of river channel geometries and subsequent training of neural networks was already performed by \cite{laloy2018training}: the resulting networks are available online, and produce rudimentary maps that depict possible geometries of river channels in a background medium, parameterised by these two binary facies. We additionally introduce prior information about marine parasequence structures created by a geological process forward model called SedSimple \citep{tetzlaffsedsimple}. SedSimple simulates sedimentary deposition, erosion and transport over geological timescales given a base topography and relative sea level curve, to create a three-dimensional geological conceptual model simulation of the subsurface. Compared to commercially available process models such as GPM \citep{otoo2021porosity}, SedSimple requires less computational resource making it possible to run a large number of simulations, but at the cost of reduced complexity in the processes and hence in the produced simulations. We train neural networks to represent the information in a large set of geometries obtained from SedSimple simulations, to produce networks that mirror those of \cite{laloy2018training}.

Parameterising the geological prior information using neural networks is important because geological process forward models cannot be used in inversion schemes directly. Their relationships between simulated geological geometries and controlling dynamic parameters and base (initial) topographies are typically both chaotically complex, and very high dimensional. This makes it very expensive, if not practically impossible, to find a set of dynamic and topographic controls that produce geometries which fit observed data. Re-parameterising the geological simulations into a more convenient (neural network based) form allows us to find geometries similar in character to those produced by the GPM, which also fit the geophysical data.

More generally, geological data are commonly available as examples (called statistical \textit{samples}) of a conceptual model. These might be facies maps from geological cross-sections, from field outcrops, or from geological process model simulations, and each of these might require a specific set of physical processes (a conceptual model) to be invoked to explain their geological origin. However, while the true parameter matrix or image in our tomographic volume may be explained using the same conceptual model, it will never exactly match one of those observed or simulated samples. Therefore, the parameterisation method must be able to generate other geological cross-sections or three-dimensional facies maps that conform to the same concept, in other words which are similar but not identical to the given set of samples. In addition, our goal is to explore the space of possible subsurface geometries to find those that are consistent with the observed geophysical data. However, exploring high-dimensional parameter spaces is extraordinarily computationally demanding, a phenomenon referred to as the curse of dimensionality \citep{curtis2001prior}. In order to make this feasible we must represent the geological information using fewer representative parameters, usually called \textit{latent} parameters. In principle we expect that this is possible because geological facies maps are not fully independent (they are spatially strongly correlated as observed in all geological outcrops -- \cite{arnold2019uncertainty}) and so may be supported by a lower-dimensional manifold of latent parameters \citep{arjovsky2017wasserstein}. Two common mathematical constructs that can be used are Variational Auto Encoders (VAEs) and Generative Adversarial Networks (GANs) -- both being types of Neural Networks. In this research, we choose GANs for their demonstrated generational quality over VAEs and the near-instantaneous generation of samples which enables more rapid inversions in our applications \citep{goodfellow2016nips}.

We invert seismic arrival times for the latent parameter posterior distribution using two methods: Markov- chain Monte Carlo (McMC) and Mixture Density Networks (MDN). The former method is computationally expensive but tends towards the correct solution in the limit of infinite sampling \citep{mosegaard1995monte}. The latter method is another neural network method which has been used succesfully to obtain marginal posterior distributions in a travel time tomography setting \citep{earp2020probabilistic, earp2020grane}. Post training, the MDN produces an estimate of the posterior distribution in near-real time, and when combined with the GAN described above, we produce a method that rapidly inverts new data to estimate solutions of the Bayesian nonlinear tomography problems that include geological prior information.

The importance of selecting \textit{appropriate} prior information is highlighted by \cite{kass1996selection}. Previous methods for including geological information in prior distributions include using a Multi-Point statistical method to simulate geology \citep{gonzalez2008seismic, lochbuhler2015summary}, Hidden Markov Models \citep{feng2018reservoir, moja2019bayesian}, and more recently using Neural Networks \citep{laloy2018training, mosser2020stochastic, song2021bridging, song2021gansim}. In this paper we also analyse cases where inappropriate geological prior information is imposed on the problem, and demonstrate that in principle such cases can be detected and corrected.

In subsequent sections we introduce our methodology in six sub-sections: first travel time tomography, then Generative Adversarial Networks, followed by Markov-Chain Monte Carlo, and Mixture Density Networks, then posterior classification probabilities, and lastly geological information. Thereafter we will describe our specific worked example, followed by a presentation of the results, a discussion, and a summary of our conclusions.
\section{Methodology} \label{chap::methodology}
\subsection{Travel time tomography}
The time that energy takes to travel between two points in a medium contains information about the part of the medium through which it propagated. In seismic or acoustic tomography the travel time stores information about variations in wave slowness (the reciprocal of wave speed) averaged over the propagation path. If multiple energy source and receiver locations are used, each travel time corresponds to a different path. In seismic tomography we use the different travel times to estimate the spatial distribution of slowness or velocity in the Earth's subsurface \citep{aki1976determination}. 

The likelihood $\rho(\bm{d}|\bm{m})$ in Equation \ref{eq:bayes_theorem} compares the travel times that would occur through a proposed parameter matrix to the observed travel times. We therefore need to compute synthetic travel times from any proposed parameter matrix, which we achieve by solving the Eikonal equation
\begin{equation}
    (\nabla t)^2=s^2
    \label{eq:eikonal}
\end{equation}
with $s(x)$ the medium slowness and $t(x)$ the arrival time from a fixed source location at any other location $x$. Equation \ref{eq:eikonal} can be solved efficiently using a finite difference approximation \citep{rawlinson2004multiple, podvin1991}, where a finer discretization of the simulation provides more accurate results. 

The comparison between the travel times of the proposed parameter matrix and the observed travel times allows a gradient direction to be computed that should infinitesimally improve the data fit. In linearised travel time tomography we iteratively update the slowness or velocity parameters by a small perturbation in that direction until a suitable data fit is achieved. Unfortunately, using that approach it is never clear whether an approximately correct parameter matrix has then been found due to the extensive and complex minima in the data misfit function. Therefore in MDN tomography we train a neural network to estimate directly the distribution of all slowness or velocity parameters that fit the data to within their uncertainties -- as explained below.

\subsection{Generative Adversarial Networks}
\begin{figure}
    \centering
    \includegraphics[width=\textwidth]{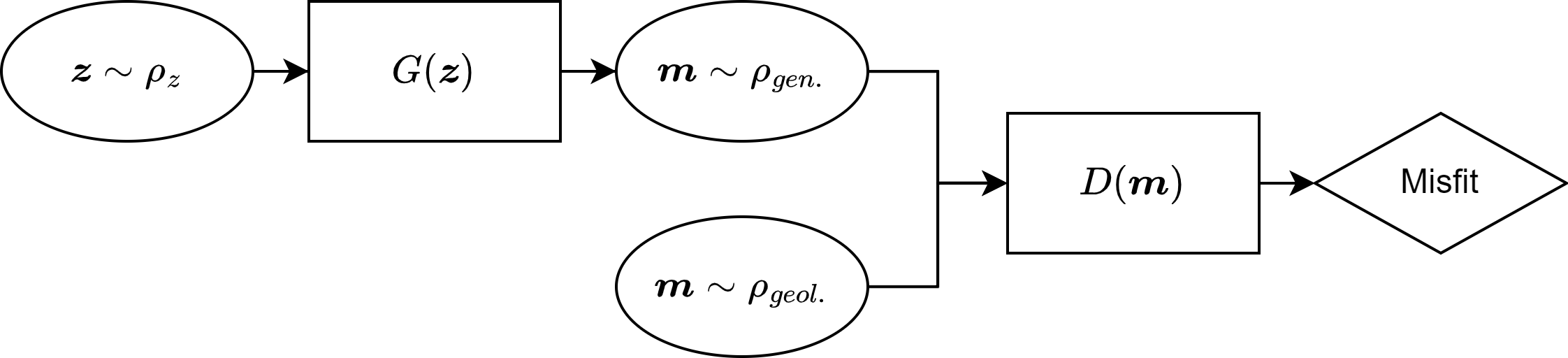}
    \caption{Schematic representation of a Generative Adversarial Network. The discriminator $D$ and generator $G$ are two independent Neural Networks. $G$ is trained to map a latent sample $\bs{z}$ to a generated simulation $\bs{m}\sim\rho_{gen.}$. $D$ is trained to discriminate between simulations coming from the generator $\bs{m}\sim\rho_{gen.}$ and geological simulations coming from the training set $\bs{m}\sim\rho_{geol.}$.}
    \label{fig:GANschematic}
\end{figure}

We store prior information about the geological concepts inside a Neural Network, specifically a Generative Adversarial Network (GAN). GANs were introduced by \cite{goodfellow2014generative} to generate high-dimensional samples efficiently from a relatively low-dimensional space of so-called \textit{latent} parameters. A GAN consists of two separate NNs: a generator and a discriminator as shown in Figure \ref{fig:GANschematic}. We train the generator to generate high-dimensional simulations that approximate simulations from a training distribution (represented by a set of samples from that distribution called the \textit{training set}). We train the discriminator to discriminate between simulations coming from the training distribution and simulations created by the generator. The training is adversarial in the sense that the discriminator is trained to minimise a loss function while the generator is trained to maximise the same loss,  so that the output of the generator approaches the training distribution, at which point the discriminator can no longer discriminate between the two distributions. After training, the discriminator is discarded and the generator is used as an efficient mapping from the lower-dimensional latent space to a higher-dimensional simulation space. In this study the high-dimensional simulations will be a geological parameter matrix $\bs{m}$ in the inversion and we will use two specific variations in the type of GAN to represent the generator: the Spatial-GAN \citep{jetchev2016texture} and the Wasserstein-GAN \citep{arjovsky2017wasserstein}.

The input to the generator $G$ is a latent vector $\bs{z}\sim\rho_{\bs{z}}$ where $\rho_{\bs{z}}$ is a distribution that can be chosen such that samples can easily be drawn; typically a Uniform or Gaussian distribution is chosen. The number of dimensions in $\rho_{\bs{z}}$ has to be large enough to describe the low dimensional manifold of the training distribution. Unfortunately the dimensionality of this manifold is unknown so a suitable number of dimensions in $\bs{z}$ is found by trial and error. $G$ can be thought of as a mapping from the latent space to the high-dimensional generated parameter matrix space, denoted as $G(\bs{z}; \theta_g)$ with $\theta_g$ representing the network parameters; this mapping is referred to simply as $G(\bs{z})$ for brevity from hereon.

The input to the discriminator is a sample $\bs{z} \sim \rho_{\bs{z}}$ randomly selected from the generator $G(\bs{z})$ or the training distribution $\bs{m}\sim \rho_{\text{geol.}}$. The discriminator outputs a scalar representing the probability that $\bs{m}$ is a sample from the geological distribution (the training set). We train $D$ to maximise the output probability of $D(\bs{m}\sim\rho_{\text{geol.}})$, and minimise the output probability of $D\left( G\left(\bs{z}\sim\rho_{\bs{z}}\right)\right)$. Conversely, $G$ is trained with the opposite goal: to maximise $\log{[1-D(G(\bs{z}))]}$. The loss for the GAN as a whole can be described by the value function $V(G,D)$:
\begin{equation}
    \min_G\max_D V(D,G)=\mathbb{E}_{\bs{m}\sim p_{\text{geol.}}(\bs{m})} \log{[D(\bs{m})]}+\mathbb{E}_{\bs{z}\sim p_{\bs{z}}(\bs{z})}\log{[1-D(G(\bs{z}))]}
\end{equation}

We do not directly optimise for $V(G,D)$ but rather update the two networks separately by alternately minimizing the following loss functions for $D$ and $G$ respectively:
\begin{align}
    \mathcal{L}_D & = -\mathbb{E}_{\bs{m}\sim\rho_{\text{geol}(\bs{m})}} \log{[D(\bs{m})]}- \mathbb{E}_{\bs{z}\sim\rho_{\bs{z}}(\bs{z})}\log{[1-D(G(\bs{z}))]} 
    \label{eq:L_D_GAN} \\
    \mathcal{L}_G & = \mathbb{E}_{\bs{z}\sim\rho_{\bs{z}}(\bs{z})}\log{[1-D(G(\bs{z}))]}
    \label{eq:L_G_GAN}
\end{align}
The adversarial objectives of the two networks implies that we seek an equilibrium between the two. Unfortunately, training to find equilibria is difficult \citep{salimans2016improved}. There are optimizers that find equilibria but none are available for the non-convex cost functions and the continuous and high-dimensional parameter spaces that occur when training GANs. Instead, a gradient descent algorithm is used that finds the low value of a cost function, and by alternating between updating $G$ and $D$ a pseudo-equilibrium is found. Nevertheless, since the optimization algorithm is sub-optimal the GAN may fail to converge during training, and it is common for multiple GANs to be trained in order to find one which performs well.

Convergence during training may be promoted by using a different loss function for $D$ that measures the distance between two distributions: $\rho_{\text{geol}}$ and $\rho_{\text{gen.}}$. \cite{arjovsky2017wasserstein} find that the Jensen-Shannon distance used in Equation \ref{eq:L_D_GAN} may not provide a gradient towards the solution in all scenarios, as it is not always differentiable. Updating Equation \ref{eq:L_D_GAN} to use the so-called Wasserstein distance measure can mitigate this problem. $\mathcal{L}_D$ then becomes
\begin{equation}
    \mathcal{L}_D = -\mathbb{E}_{\bs{m}\sim\rho_{\text{geol.}}} D(\bs{m}) + \mathbb{E}_{\bs{z}\sim\rho_{\bs{z}}} D(G(\bs{z}))
\end{equation}
The Wasserstein distance is shown to be continuous and differentiable almost everywhere \citep{arjovsky2017wasserstein}. The improvements to GAN training include improved training stability and reduced mode collapse (the latter is the term used to describe situations where multiple high-dimensional features are mapped to the same latent parameter by the GAN, thus restricting its generation capability). 

\cite{jetchev2016texture} update the GAN architecture to make it better suited to synthesizing textures or maps; the updated architecture is called a Spatial-GAN or SGAN. Texture synthesis is the generation of samples of a given texture, which is defined as repeating patterns with some degree of stochasticity \citep{georgiadis2013texture}. Geological parameter matrices can be similar to textures as they often contain approximately repeating patterns and have some stochasticity (e.g., repeating sedimentary layers with varying thicknesses, or meandering river channels consisting of similar facies). The input to an SGAN is a latent tensor (matrix) rather than simply a latent vector as used in a standard GAN. Furthermore, all the layers in the SGAN are convolutional layers \citep{jetchev2016texture}. This enables us to scale the input tensor to obtain a different sized output, i.e., in our case a larger latent tensor will result in a larger geological parameter matrix. What is more, individual elements of the latent tensor describe only a single, localised patch of the output parameter matrix. Thus, we can update a single patch of an output parameter matrix while keeping the rest of the matrix constant. The discriminator is also updated to output a loss for each entry of the input tensor. 

\subsection{Markov-chain Monte Carlo}
We would like to obtain a reasonable accurate estimate of the posterior distribution $\rho(\bs{m}|\bs{d})$ in Equation \ref{eq:bayes_theorem}. We could sample prior distribution $\rho(\bs{m})$ and compute the likelihood for those samples to obtain an estimate of the posterior distribution. However, the maxima in $\rho(\bs{m})$ do not necessarily align with the maxima in $\rho(\bs{m}|\bs{d})$ which makes sampling of the posterior inefficient. What is more, the likelihood $\rho(\bs{d}|\bs{m})$ may introduce nonlinear relationships that further impede finding a representative sampling of $\rho(\bs{m}|\bs{d})$. We would therefore like to sample $\rho(\bs{m}|\bs{d})$ directly, which is approximately possible using Markov-chain Monte Carlo (McMC) sampling \citep{mosegaard1995monte}. McMC sampling creates a chain of samples, where each sample is found using a two step process: first, we sample a parameter according to a proposal distribution $q(\bs{m'}|\bs{m})$ i.e., the probability of moving from a current sample $\bs{m}$ to a proposed sample $\bs{m'}$. Second, the proposed sample is accepted or rejected depending on the misfit $S(\bs{m})$ between the data generated from the parameter sample and the observed data. The probability of acceptance
\begin{equation}
    P_{accept}=\begin{cases}
    1 & \text{if } S(\bs{m'}) \le S(\bs{m}) \\
    \exp{-\frac{\Delta S}{\sigma^2}} & \text{if } S(\bs{m'}) > S(\bs{m})
    \end{cases}
    \label{eq:mcmc_acceptance}
\end{equation}
where $\Delta S = S(\bs{m'}) - S(\bs{m})$ and we have assumed an explicit form for the likelihood $\exp{-\frac{\Delta S}{\sigma^2}}$, where $\sigma^2$ is the variance or noise on the observed data. Thus, if the misfit of the proposed parameter $\bs{m'}$ is lower than the current parameter $\bs{m}$ then $\bs{m'}$ is always accepted as being a sample of the posterior. In the reverse case, $\bs{m'}$ is accepted as a posterior sample with a probability based on the difference between the misfits of the proposed and current model if $\bs{m'}$ is rejected, the current model is repeated (duplicated) in the chain. 

From an initial parameter sample, consecutive samples are found using Equation \ref{eq:mcmc_acceptance}. \cite{mosegaard1995monte} show that after an infinite number of samples, the set of parameter samples is distributed according to the posterior distribution. However, an infinite number of samples is computationally infeasible, as is a number of samples that is sufficiently large to approximate this case. Therefore multiple chains are computed in parallel, each with different initial parameters, such that we obtain a greater diversity of samples more rapidly. What is more, having multiple chains allows for more resilience if a chain gets stuck in a maximum. 

Although the acceptance probability in Equation \ref{eq:mcmc_acceptance} ensures that we end up with a set of samples that estimate the posterior distribution if sampled into infinity, we would like a finite set of samples that are representative of the posterior distribution. What is more, we would like to obtain this finite set efficiently. Therefore, we must design our proposal distribution such that we minimise the number of rejected samples while still spanning the parameter space. We define our proposal distribution as randomly selected perturbation to parameters of the previous parameter matrix in the chain. We can vary both the quantity of the perturbation and the number of parameters updated to optimise the efficiency of posterior distribution sampling. Even after such provisions the samples are only ever approximately distributed according to the posterior. We therefore often only analyse moments of the sample set (means, variances, etc.). These are assumed to be more robust estimators of properties of the posterior distribution than are the individual samples.

\subsection{Mixture Density Networks}
\label{chap:MDN}
Mixture Density Networks (MDN) are a type of Neural Network that can be trained to infer the posterior distribution of parameters from measured data \citep{bishop2006pattern}. The posterior  $\rho(\bs{m}|\bs{d})$ is approximated by a weighted sum of multiple Gaussians 
\begin{equation}
    \rho(\bs{m}|\bs{d}) = \sum^K_{k=1}{\pi_k(\bs{d})\NormalDist\left(\bs{m}|\mu_k(\bs{d}),\sigma^2_k(\bs{d})\right)}
\end{equation}
where $K$ is the number of Guassians, $\pi_k$ the $k$th Gaussian's mixing coefficient or weight, $\NormalDist$ is the Gaussian or Normal distribution, $\mu_k$ the Gaussian mean and $\sigma^2_k$ the Gaussian standard deviation. The vector parameters $\bs{\pi}$, $\bs{\mu}$, and $\bs{\sigma}$ are inferred from the data using a neural network which may have any of a range of architectures and complexities.

MDN training is performed with $N$ parameter-data pairs $\{(\bs{m}_n, \bs{d}_n):n=1, \dots,N\}$, which are generated by selecting $\bs{m}_n$ according to the prior pdf and calculating the corresponding measured data $\bs{d}_n$ by using a synthetic forward model. The neural network weights $\theta_{\text{MDN}}$ are optimised by minimising a cost function $E$ which for independent training data is taken to be
\begin{equation}
    E(\theta_{\text{MDN}})=-\sum^N_{n=1}{\ln{\left[\sum^K_{k=1}{\pi_k(\bs{d}_n,\theta_{\text{MDN}})\NormalDist(\bs{m}|\mu_k(\bs{d}_n, \theta_{\text{MDN}}),\sigma^2_k(\bs{d}_n, \theta_{\text{MDN}})))}\right]}}
    \label{eq:mdn_loss}
\end{equation}
with $N$ the number of pairs in the training set \citep{bishop2006pattern}. To optimise the network, we calculate derivatives of the cost function with respect to each network weight, which are obtained using a back propagation procedure \citep{bishop2006pattern}. Due to the sum over all data points in Equation \ref{eq:mdn_loss} we back propagate the derivatives for each data point then sum the resulting $N$ derivatives to find the derivative of $E(\theta_{\text{MDN}})$.

\subsection{Posterior Classification Probabilities}
A neural network tasked with classification of its inputs into a set of discrete classes often outputs a score for each possible class. The classification derived by the network is then the class with the greatest score. It turns out that if the network is trained with specific cost functions that the outputs for each class are equal to the Bayesian posterior probabilities \citep{richard1991neural}, provided that the network is trained using a one-hot encoding scheme (i.e., the true classification is encoded as $1$ for the correct class and $0$ for the other classes) and that the network has a sufficient number of trainable weights. \cite{richard1991neural} prove this for a squared-error and a cross-entropy cost function. 

Assume we have a training set containing a parameter matrix $\bs{m}$ which belongs to one of $N$ classes in $\{C_n:n=1,\dots,N\}$. Let $\{y_n:n=1,\dots,N\}$ be the network output and $\{c_n:n=1,\dots,N\}$ the desired output. Then, we can construct the squared error cost function
\begin{equation}
    E(\theta_{class})=\mathbb{E}\left\{\sum^{N}_{n=1}[y_n(\bs{m})-c_n]^2\right\}
    \label{eq:classifier_loss}
\end{equation}
with $\theta_{class}$ the network weights, $y_n(\bs{m})$ the network output for class $n$, and $c_n$ the desired output for class $n$ \citep{richard1991neural}. The neural network weights are then optimised in a similar fashion to the MDN (see section \ref{chap:MDN}).

\subsection{Geological Information}
The conceptual geological model usually describes our beliefs about the tectonic setting, depositional environment of sediments, geographical relationships to continents and marine waters, and other high level information. This model ultimately governs lower level information about the exact geometry, abundance of different sediment types, et cetera. We use two conceptual models: a braided river system and marine parasequences. The former model has already been described in detail in \cite{laloy2018training}, so here we discuss a relatively new marine parasequences conceptual model.

Our choice of conceptual model defines what processes are included in the SedSimple \citep{tetzlaffsedsimple} geological process modeller (GPM) runs used herein. A GPM simulates geological processes through time computationally to obtain a 3-dimensional distribution of geological facies such as that shown in Figure \ref{fig:gpm3d} \citep{tetzlaff1989simulating}. Simulations are started from an initial topography, for which in this study we use a sigmoidal topography that represents the transition from the continental shelf to the abyssal plain. Relative sea level through time is defined to be sinusoidal and we include an influx of water and sediment on the shelf representing a river. Last, we define two types of clastic sediments by setting their relative transportability; one sediment has double the transportability of the other. Note that in this manuscript we aim to establish and demonstrate the methodology rather than to apply it in a real setting. We therefore chose these values to obtain a thick sediment with interesting features rather than to emulate a scenario matching a certain geographical location and time interval as would be the case when this method was deployed in practise. All parameter values used herein are defined in Table \ref{tab:gpm_param}.

\begin{table}
\caption{Parameters used for the geological process model simulation shown in Figure \ref{fig:gpm3d}. Transportability in the table is also known as the diffusion coefficient.}
\label{tab:gpm_param}
\centering
\begin{tabular}{l|l}
Parameter                   & Value \\ \hline
Geological time             & 500 kA \\
Manning coefficient         & 0.3   \\
Water source                & 1,000 m$^3$/s \\
Sediment 1 influx           & 60 mL/s  \\
Sediment 2 influx           & 160 mL/s \\
Transportability Sediment 1 & 100  \\
Transportability Sediment 2 & 200 
\end{tabular}
\end{table}

After the initial parameters are defined, SedSimple simulates different geological processes at each of a set of small time steps. Such processes are: sedimentary dispersion, erosion, transportation and deposition. Dispersion is the process that simulates sediment particles moving downhill from high to low elevations, and similarly from high to low concentrations when suspended in a fluid. The governing equation for sediment dispersion is 
\begin{equation}
    \frac{\partial z}{\partial t}=D\nabla^2z+s
\end{equation}
with elevation $z$, time $t$, diffusion coefficient $D$, and the sediment source term $s$. Sedimentary erosion, transport and deposition are dependent on the fluid flow. SedSimple simulates fluid flow to determine whether sediment is eroded, transported, and deposited, for which a simplified version of the Navier-Stokes equation is employed. The erosion and deposition are then calculated based on empirical formulae (see \cite{tetzlaff1989simulating} for more information). 

Although GPM algorithms are deterministic they are still chaotically nonlinear. A small change in the input variables can cause large changes in the output simulation. For example, if the initial simulation has two parallel river channels then minute changes in the input parameters will create entirely different braided river systems. It is therefore difficult, if not impossible, to use a GPM directly in an inversion algorithm to fit specific measured data. We therefore train a GAN to emulate the spatial sedimentary patterns produced by the GPM \citep{mosser2020stochastic}. Since the generator in a GAN has an analytic structure, the input latent paramaters can be varied so that the distribution of simulations produced emulates the prior distribution represented by any given set of simulations from the GPM. This method allows multiple GPM simulations with different initial parameters to be used to capture the chaotic nature of sedimentary distributions.
\section{Problem Description} \label{chap::problem_description}
Our goal is to focus on how we introduce geological information into tomographic studies, in particular in the case where we may have multiple conceptual geological models. Given that we can deploy two such models (one parameterised by \cite{laloy2018training}, and one using SedSimple), we can investigate the impact on geophysical tomography if different (and potentially misleading) prior information is injected.

\begin{figure}
    \centering
    \includegraphics[width=\textwidth]{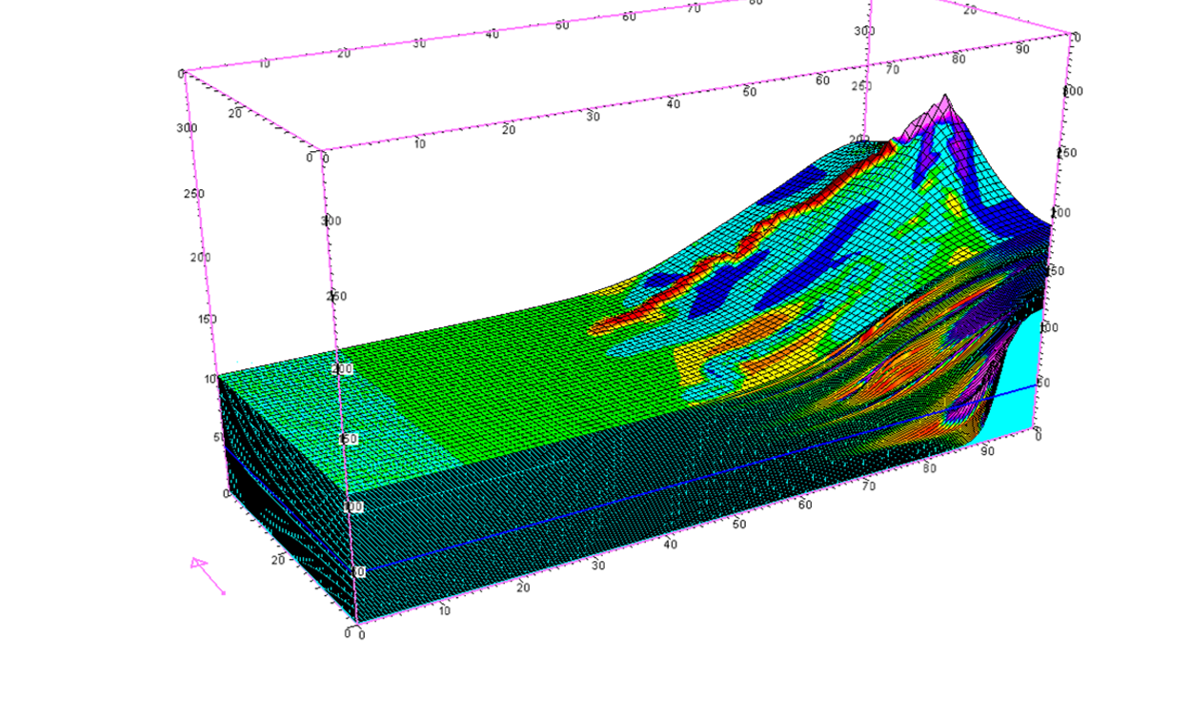}
    \caption{3D view of an example output from the geological process model SedSimple. Colours in the plot represent the relative concentrations of different facies in the simulation.}
    \label{fig:gpm3d}
\end{figure}

The SedSimple simulation in Figure \ref{fig:gpm3d} was used to construct training data for the GANs that embody the second conceptual model. In this simulation there are two different sediments, both siliciclastics but with different grain sizes. The colours in the plot represent the abundance of the two sediments. We took 1,800 two-dimensional 32-by-32 pixel slices in both the $x$ and $y$ direction from the 3-dimensional simulation which created a training set of parameter matrices for the GAN. For details of the braided river system conceptual model and SGAN training we refer the reader to \cite{laloy2018training}. 

We trained a GAN to emulate the marine parasequences conceptual model using codes from \cite{kang2020ContraGAN}. We tested different network architectures including BigGAN \citep{brock2018large}, ReACGAN \citep{kang2021ReACGAN}, ICRGAN \citep{zhao2020improved}, and WGAN-GP \citep{gulrajani2017improved}. For each network we minimised the number of latent parameters by visual inspection of the output, and trained the same architecture multiple times to reduce the effect of poor (random) network initialisation. The trained networks are visually compared after which we selected a single best network, in our case a WGAN with 8 latent parameters. Training of this network took 7 hours and 30 minutes on a single NVIDIA TITAN X. 

\begin{figure}
    \centering
    \begin{subfigure}{0.49\textwidth}
        \centering
        \includegraphics[width=0.9\linewidth]{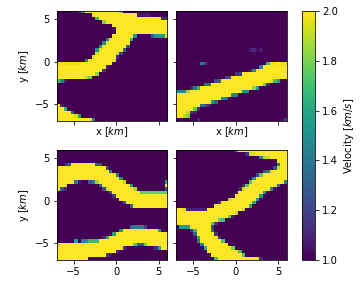}
    \end{subfigure}
    \begin{subfigure}{0.49\textwidth}
        \centering
        \includegraphics[width=0.9\linewidth]{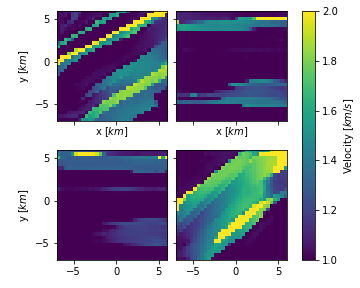}
    \end{subfigure}
    \caption{Four realisations from the braided river system (left) and marine parasequences (right) GANs. Colours represent seismic velocities.}
    \label{fig:gan_samples}
\end{figure}

The parameter matrices generated by both the braided river and the marine parasequence GAN are in a value range of $[-1, 1]$ which we rescaled to $[1, 2]~km/s$ to represent a reasonable range of seismic velocities. Four samples from each GAN are shown in Figure \ref{fig:gan_samples}: the braided river channel realisations on the left and marine parasequences on the right represent samples from the two prior pdf's used in this study. Although the parameter matrices are high-dimensional, there is a low-dimensional latent representation for each such matrix. The braided river channel prior is encoded by 9 latent parameters and the marine parasequences prior is encoded by 8 latent parameters. The respective GANs create mappings from the low-dimensional latent parameters to the prior pdf's in high-dimensional geophysical parameter matrices. Each set of latent parameter values selected from the latent distributions produces an approximate sample from the corresponding geological prior pdf.

\begin{figure}
    \centering
    \begin{subfigure}{0.49\textwidth}
        \centering
        \includegraphics[width=\linewidth]{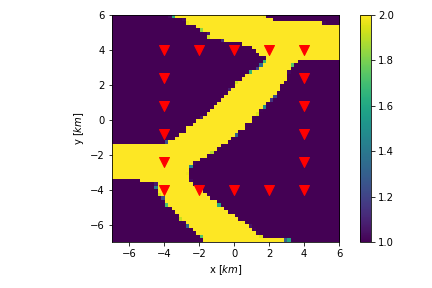}
    \end{subfigure}
    \begin{subfigure}{0.49\textwidth}
        \centering
        \includegraphics[width=\linewidth]{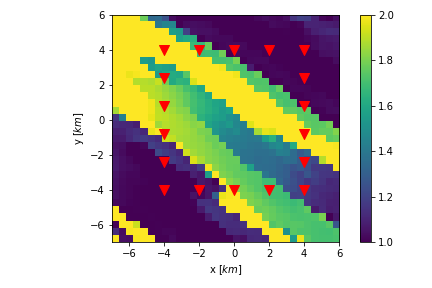}
    \end{subfigure}
    \caption{The two 'true' parameter matrices for which we simulate measured data, one from each of the conceptual models: a braided river system (left) and marine parasequences (right). Red triangles indicate the data acquisition geometry -- the locations of both sources and receivers between which travel time data are simulated.}
    \label{fig:true_models}
\end{figure}

To represent a geophysical tomographic survey we defined a square data acquisition geometry with corners at $x=-4,4km$ and $y=-4,4km$ and a source spacing of $1.4km$ as shown by the red triangles in Figure \ref{fig:true_models}. This geometry defines the locations of sources and receivers: each location in turn acts as a seismic source while all other locations act as receivers. This generated $153$ unique travel times -- the data corresponding to each parameter matrix. 

Figure \ref{fig:true_models} shows what we used as the `true' parameter matrices composed of a terrestrial river system (left) and marine parasequences (right). The acquisition geometry is indicated by red triangles. In the examples below, we inverted the travel times corresponding to these true parameter matrices for the Bayesian posterior pdf in the low-dimensional latent space in each case. Samples of this pdf could then be translated to the high-dimensional parameter space using the GAN. This indirect approach ensures that we need only estimate the posterior pdf over 8 or 9 latent parameters rather than across the high-dimensional 32-by-32 parameter matrix.

For each data set we computed two estimates of the posterior pdf: a benchmark solution found using McMC and a large number of samples, and a rapid estimate using MDNs. We validated the convergence of the McMC runs by monitoring the posterior marginal pdf estimates for each parameter in latent space. Any chains that were obviously stuck in local minima (span a relatively narrow range of parameter values) were removed and we validated that a reasonable number of samples have been taken by ensuring that the posterior distribution is essentially the same for the complete set as well as half of the set of posterior samples. We computed 40 chains with around two million samples each which took approximately 3 days to run. This is a lot of samples for McMC runs with $\sim$8 parameters, but the latent parameters are more information dense compared to model parameters and can be strongly multi-modal, so we run many samples to avoid remaining trapped in a subset of the modes.

For each prior pdf we trained an MDN to invert the data using around 3 million samples in each training set. During training we monitored progress by evaluating a validation dataset of 25\% of the size of the training set. Multiple network architectures were used and each architecture is trained multiple times to eliminate bias due to their random initialisation. Using a test set of size 5\% of the training set size we measured the network performance. To better generalise the outputs, we selected 5 networks and combined them in a mixture of experts. We found that training the MDN for the complete posterior resulted in the MDN posterior estimate not finding all the modes present in the `true' McMC posterior distribution. We therefore opted to train multiple networks to infer the marginal pdf of a single latent parameter, similarly to the setup in \cite{earp2020probabilistic}. Thus, for a single prior we have $N_{\text{latent parameters}}\times M_{\text{mixture of experts}}$ neural networks (e.g., the river channel prior pdf has $9$ latent parameters giving $9\times5=45$ networks). Training of a single network took about 90 minutes using a single NVIDIA Tesla K80. We had access to multiple GPUs so we could train the networks in parallel.

Finally, we trained a classifier NN that estimates the posterior probability that each prior pdf pertained to a certain set of travel times. For this we combined the river and marine training sets such that we obtained a total of 6 million data points. We trained multiple networks with different architectures and selected the one that performs best on a validation dataset. We did not use the complete training data set but instead randomly selected a number of samples for each training epoch. For the optimal network we used 8 batches with 4,000 samples each and trained the network for 100 epochs. This took 80 seconds on a NVIDIA Tesla T4.
\section{Results}\label{chap::results}
\begin{figure}
    \centering
    \includegraphics[width=\textwidth]{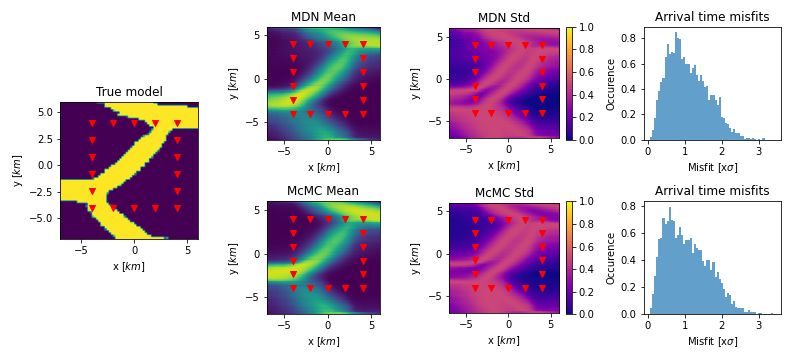}
    \caption{The true braided river channels parameter matrix (left) and the summary statistics for the posterior estimates from the MDN (top) and McMC (bottom). The statistics are the posterior mean, posterior standard deviation, and the travel time misfits.}
    \label{fig:results_river}
\end{figure}

Figure \ref{fig:results_river} shows the estimated posterior statistics for a river channel true parameter matrix where the inversion is performed using the correct prior. From left to right the figure shows: the true parameter matrix, the posterior mean for MDN (top) and McMC (bottom), the posterior estimate standard deviation for MDN (top) and McMC (bottom), and lastly a histogram of the arrival time misfits for 5,000 posterior samples from MDN (top) and McMC (bottom) solutions (the MDN samples are independently selected from the posterior marginal pdf of each latent parameter). The means for both posterior estimates are close to the true parameter matrix. The standard deviations show high uncertainty loops (boundaries) around features in the true parameter matrix \citep{galetti2015uncertainty} similar to \cite{earp2020probabilistic}; these are expected, and quantify uncertainty in the location of the edges of those features. The posterior estimates thus show that there is a river channel running diagonally over the parameter matrix but that the exact boundary of the channel is uncertain. The channels outside of the survey acquisition area have broader uncertainties since fewer rays travel outside of the acquisition array. The travel time misfits are centred around one travel time measurement error $\sigma$ and almost all are within two measurement errors from the true arrival times. Finally, all statistics shown here are consistent between our benchmark McMC posterior and the rapid MDN method. All of the above indicates that both the McMC and MDN solutions are approximately correct, and conform to intuition about probabilistic solutions derived from analyses in previous studies.

\begin{figure}
    \centering
    \includegraphics[width=\textwidth]{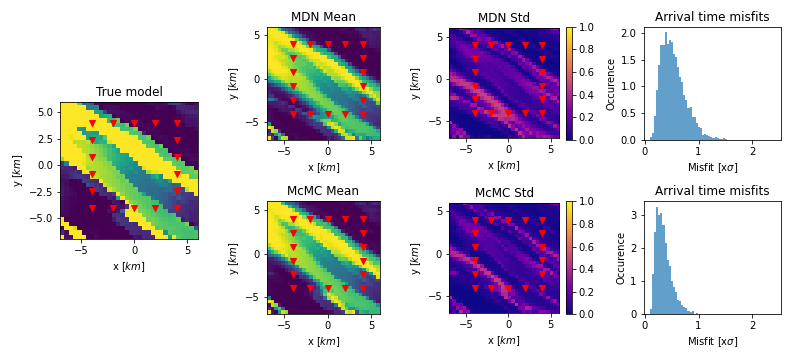}
    \caption{The true marine parasequences parameter matrix (left) and the summary statistics for the posterior estimates from the MDN (top) and McMC (bottom). The statistics are the posterior mean, posterior standard deviation, and the travel time misfits.}
    \label{fig:results_marine}
\end{figure}

Figure \ref{fig:results_marine} shows similar plots to Figure \ref{fig:results_river} but for the marine parasequence true parameter matrix and prior information. Again, there is a close match between the posterior mean and the true parameter matrix. The McMC posterior does resolve the high-velocity feature at location $(1,1)km$ better than the MDN posterior mean velocities. The MDN posterior standard deviations are wider for the transition from low- to high-velocity at $(-3,-2.5)km$ compared to the McMC standard deviations. The McMC posterior estimate is thus more narrowly concentrated around the true parameter matrix. This is confirmed by the arrival time misfit histograms: the McMC posterior data misfits show a narrower peak at a lower misfit value compared to the broader peak of the MDN posterior data misfit. This illustrates that the MDN posterior marginal pdf estimates do not capture all of the information that is contained in the McMC posterior. This is likely to be because we infer only single-parameter MDN marginal distributions which therefore do not contain information about correlations between latent parameters; by contrast, the Monte Carlo samples are taken in the full latent space and so do contain correlation information. While in principle it is possible to train MDNs to represent the full correlated posterior pdf, we found such networks extremely difficult to train reliably. Therefore, this reduction in  posterior information is the price paid for obtaining stable solutions for any travel time data set in $\sim1s$ from an MDN rather than from days of run time when using McMC.

\begin{figure}
    \begin{subfigure}{0.49\textwidth}
    \centering
    \includegraphics[width=\linewidth]{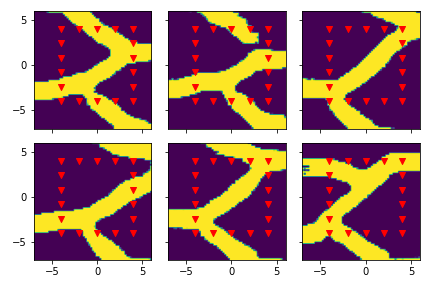}
    \caption{River true model inversion}
    \label{fig:posterior_river_samples}
    \end{subfigure}
    \begin{subfigure}{0.49\textwidth}
    \centering
    \includegraphics[width=\linewidth]{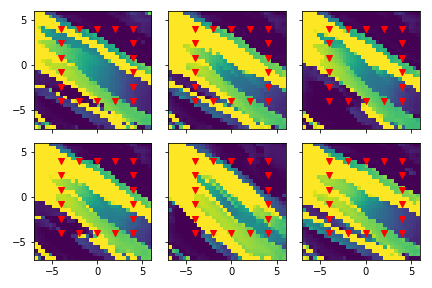}
    \caption{Marine parasequences true model inversion}
    \label{fig:posterior_marine_samples}
    \end{subfigure}
    \caption{Samples from the MDN posterior marginal pdf estimates. Left shows samples from the inversion in Figure \ref{fig:results_river}, right shows samples from the inversion in Figure \ref{fig:results_marine}.}
\end{figure}

The statistics in Figure \ref{fig:results_river} and \ref{fig:results_marine} only show summary statistics of the posterior pdf. What is more, the posterior mean is not in itself a geological parameter matrix selected from the posterior distribution (the values shown are an integral over all parameter matrix samples). We therefore show six samples from the MDN posterior estimate in Figure \ref{fig:posterior_river_samples} and \ref{fig:posterior_marine_samples} for the river and marine inversion respectively. The samples in each set are slightly different but all do resemble their respective true parameter matrix. What is more, all samples are geological: the samples from the river inversion all show river channels and all samples from the marine parasequence inversion show geological marine parasequences. This is not likely to occur in any inversion conducted using a non-geological prior (e.g., \cite{bodin2009seismic, galetti2015uncertainty, earp2020probabilistic}).

\begin{figure}
    \centering
    \includegraphics[width=\textwidth]{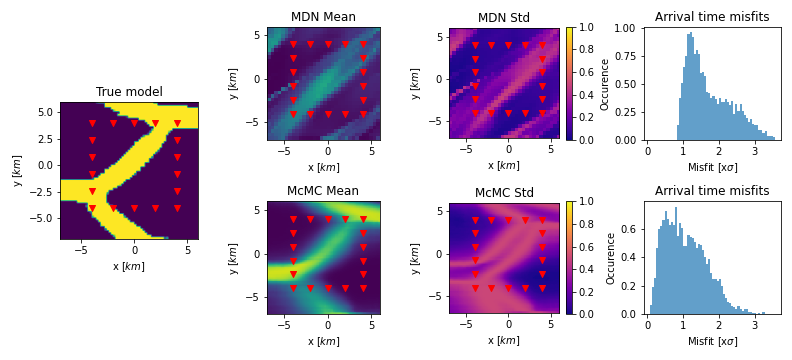}
    \caption{The true parameter matrix taken from the river prior (left) and the summary statistics for the posterior estimates using the unsuitable marine prior pdf (top) and suitable river prior pdf (bottom). The unsuitable posterior pdf estimates are computed using the MDN.}
    \label{fig:results_river_wrong}
\end{figure}

\begin{figure}
    \centering
    \includegraphics[width=\textwidth]{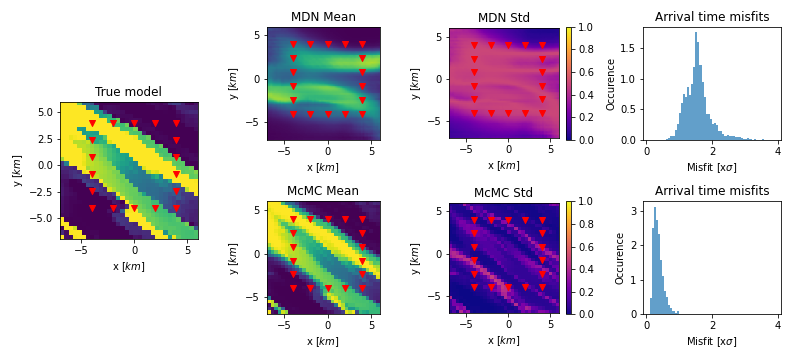}
    \caption{The true parameter matrix taken from the marine parasequences prior (left) and the summary statistics for the posterior estimates using the unsuitable river prior pdf (top) and suitable marine parasequences prior pdf (bottom). The unsuitable posterior pdf estimates are computed using the MDN.}
    \label{fig:results_marine_wrong}
\end{figure}

\begin{figure}
    \begin{subfigure}{0.49\textwidth}
    \centering
    \includegraphics[width=\linewidth]{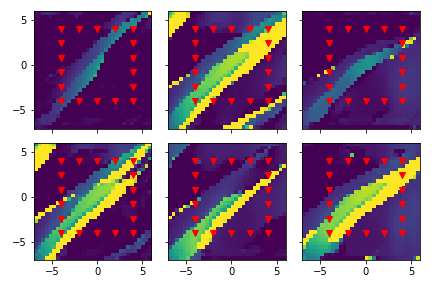}
    \caption{River true model}
    \label{fig:posterior_river_wrong_samples}
    \end{subfigure}
    \begin{subfigure}{0.49\textwidth}
    \centering
    \includegraphics[width=\linewidth]{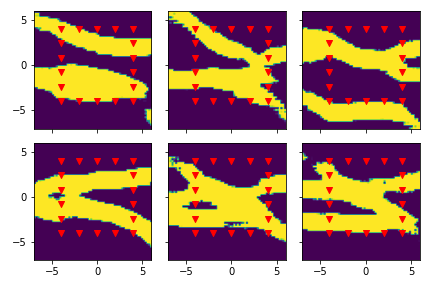}
    \caption{Marine parasequences true model}
    \label{fig:posterior_marine_wrong_samples}
    \end{subfigure}
    \caption{Samples from inversions using unsuitable prior information. Left shows the samples from the inversion in Figure \ref{fig:results_river_wrong}, right shows the samples from the inversion in Figure \ref{fig:results_marine_wrong}.}
\end{figure}

So far we have applied the correct set of prior information for each of our target true parameter matrices. However, it has been shown that prior choices between conceptual geological models \citep{bond2015structural} and their parameters \citep{curtis2004optimal} are subject to natural human biases \citep{polson2010dynamics, curtis2012science, bond2012makes}. It is therefore of interest to assess the effects of using incorrect prior information: inverting a river channel true model using a marine prior pdf and vice versa. Figures \ref{fig:results_river_wrong} and \ref{fig:results_marine_wrong} show the inversions using the incorrect prior pdf (top) versus the correct one (bottom). Figure \ref{fig:posterior_river_wrong_samples} and \ref{fig:posterior_marine_wrong_samples} show example posterior samples for the river and marine true parameter matrices, respectively. The posterior statistics show that the inversions with the correct prior perform better than those with the incorrect prior, as expected. However, the incorrect prior inversions do still retrieve reasonable estimates of the overall structure. The  inversion of the river channel data produces a diagonal channel-like feature in all samples in Figure \ref{fig:posterior_river_wrong_samples} but the channels sub-parallel to the x-axis at the top and bottom of the true model are not inferred. The posterior samples from the inversion of the marine data show that the posterior does somewhat capture the high-velocity feature in the true model but that there is a lot of variation. However, in both cases when inappropriate prior information is used, the expected loop-like uncertainty structures created by high uncertainties occurring on the boundaries of the main velocity anomalies do not occur.

\begin{figure}
    \centering
    \includegraphics[width=\textwidth]{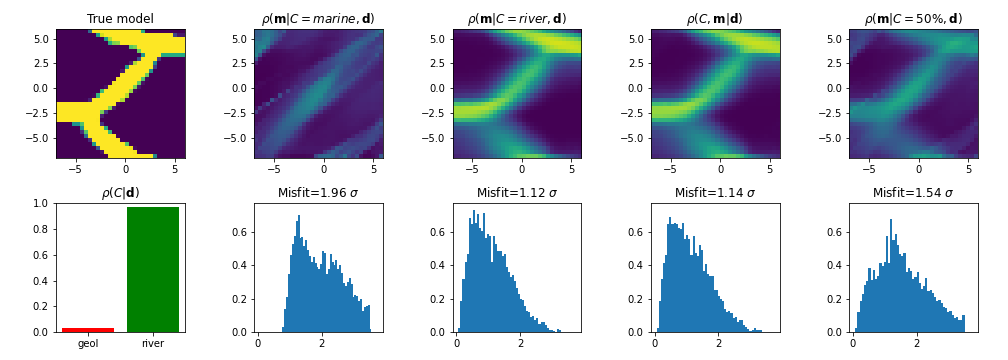}
    \caption{Posterior statistics for different prior pdfs and prior combinations. Left column shows the true river parameter matrix (top) and the posterior probability of the conceptual model class given the observed travel times (bottom). Successive columns show the posterior mean (top) and travel time misfit (bottom) for the following scenarios: marine parasequence conceptual model, river prior conceptual model, joint posterior probability of model parameters and conceptual model class, and the posterior mean if the posterior probability for each conceptual model class was Uniform.}
    \label{fig:priorclass_river}
\end{figure}

\begin{figure}
    \centering
    \includegraphics[width=\textwidth]{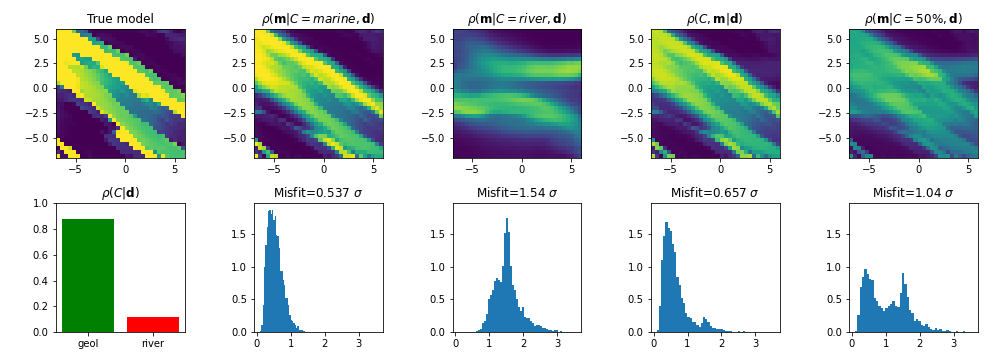}
    \caption{Posterior statistics for different prior pdfs and prior combinations. Left column shows the true marine parasequences parameter matrix (top) and the posterior probability of the conceptual model class given the observed travel times (bottom). Successive columns show the posterior mean (top) and travel time misfit (bottom) for the following scenarios: marine parasequence conceptual model, river prior conceptual model, joint posterior probability of model parameters and conceptual model class, and the posterior mean if the posterior probability for each conceptual model class was Uniform.}
    \label{fig:priorclass_marine}
\end{figure}

The data misfits in Figures \ref{fig:results_river_wrong} and \ref{fig:results_marine_wrong} highlight the difference between the correct and incorrect conceptual models. The top rows show posterior statistics obtained using unsuitable prior information and the bottom row show comparable results when using appropriate prior information. Differences between the misfit distributions (right most panels) for the same data set indicates that there may be information in the data that could discriminate which class of prior information is more likely to be appropriate. We therefore train a classifier neural network that provides the posterior class probability of each conceptual model $C$ given the observed travel times $\rho(C|\bs{d})$. Given this posterior we can also compute the joint posterior probability of the conceptual class and the corresponding model parameters $\rho(C,\bs{m}|\bs{d})$. The results are shown in Figures \ref{fig:priorclass_river} and \ref{fig:priorclass_marine} for the river and marine true parameter matrices respectively. The top panels show the following geological parameter matrices from left to right: the true parameter matrix, the posterior mean using marine prior information, the posterior mean using river prior information, the joint posterior mean of class and model parameters, and lastly a posterior mean if the class probabilities were Uniform (that is, in a scenario in which we do not know which conceptual model was more likely). The bottom row shows the posterior class probabilities on the left, and then histograms of the data misfits corresponding to the posterior pdfs in the top row. First, the posterior class probability provides clear information about which conceptual model pertains to which data set. As above, the posterior misfits are lower when the correct conceptual model is used for the inversion. If we combine the class probabilities and the parameter probabilities into the joint posterior probability $\rho(C,\bs{m}|\bs{d})$, the misfits are higher than for the correct prior but significantly lower than those obtained using either the wrong class or Uniform class probabilities.

\section{Discussion} \label{chap::discussion}
Our aim in this paper is to develop and demonstrate a methodology to inject high level (conceptual) and lower level (parameter) geological prior information into a Bayesian inversion scheme and to investigate their effects. Figures \ref{fig:results_river} and \ref{fig:results_marine} show that we can successfully inject geological prior information into Bayesian inversion using the GAN. For both the river and the marine prior pdfs we found posterior pdfs that closely match the true parameter matrix. In these cases, the rapid MDN inversions obtain a similar posterior to the McMC posterior in a fraction of the time, post-training. Furthermore, we have shown that the GAN prior parameterisation is reasonably agnostic to the Bayesian inversion scheme used (e.g., McMC or MDN) to estimate parameter pdfs. Thus, if faster or more accurate inversion methods become available our solution to injecting geological prior information can still be used.

However, the true parameter matrices that we have used in these examples result in so-called inverse crimes. An inverse crime is a situation where a parameter matrix which has been simplified to a certain degree, is used to generate data, which are then inverted to try to recover a parameter matrix that contains the same simplifications \citep{kaipio2007statistical}. Inverse crimes show a best case scenario for any inversion method since the method is not tested against the complexity of true data; such tests might overestimate the performance of an algorithm. We therefore also show inversions of the true parameter matrix using the incorrect prior conceptual model. In that scenario there is no inverse crime as the simplifications made in the true parameter matrix and in the prior information are different. Although these inversions performed worse than their inverse crime equivalents, we argue that the results are potentially still acceptable, particularly when using Monte Carlo methods. 

Another benefit of using geological prior pdfs is that the resulting posterior pdf consists only of geologically reasonable parameter matrices. This means that each sample of the posterior pdf is a geological parameter matrix in itself. Posterior samples from most tomographic methods can contain high probability parameter matrices that are not geological \citep{earp2020probabilistic}; when interpreting such non-geological samples, the interpreter always has a conundrum: whether to interpret the result as a high-probability parameter matrix that is not geologically feasible or a lower-probability parameter matrix that is geologically feasible? Neither option is ideal, and altering the high-probability parameter matrix to make it more geological adds unknown uncertainty to the parameter matrix. One option is to interrogate the whole set of parameter samples to answer specific geological questions \citep{arnold2018interrogation, zhang2022interrogating, zhao2022interrogating}. On the other hand, our method solves this problem directly, as each parameter matrix in the posterior is geological. High-probability samples can therefore be interpreted directly, and their relative probabilities are known. 

If more prior pdfs become available e.g., conceptual models for salt diapirs, carbonates, or specific geographic regions, it may become infeasible to invert the data for each of these priors due to the computational expense that comes with inverting the data. As a solution we introduced a classifier neural network that infers posterior probabilities of which prior pdf is most consistent with the arrival time data. This information can be especially useful in the case when there is a wish to obtain the most accurate possible McMC posterior estimates, but where the posterior parameter estimation is costly to compute, as the arrival times will only have to be inverted for the parameters associated with the prior information class found by the classifier NN. This neural network must be retrained for each additional prior pdf but its training is cheap relative to the cost of training the MDNs or computing an McMC posterior estimate. Furthermore, it can be used to combine different prior pdfs to obtain a posterior parameter estimate given different classes of prior information, which could be useful in a scenario where the true parameter matrix is in fact best represented by a mixture between two canonical conceptual models.

Finally, this work highlights the improvements to imaging using MDNs combined geological prior information. \cite{earp2020probabilistic} used MDNs to invert directly for parameters that represent pixels in the tomographic image, and obtained marginal posteriors for each parameter individually. By introducing the geological prior pdf we also obtain uncorrelated marginal pdfs, but on latent parameters of the GANs; varying those latent parameters within their respective posterior marginals translates through the GAN to correlated estimates of the posterior variation of the image parameters. This is demonstrated by the fact that the posterior samples look geological and are thus highly-correlated in space. This occurs because approximately correct intra-parameter correlations are stored in the GAN architecture. Ideally, the latent parameters would also be correlated a posteriori, and this may be possible in the future using different neural network architectures within the MDN, or different training methods.
\section{Conclusion} \label{chap::conclusion}
In this study we inject geological prior information into a travel time tomography Bayesian inversion scheme to improve the posterior knowledge about parameters that describe the tomographic image. We evaluate two different geological conceptual models: a braided river system and the formation of marine parasequences. Both are parameterised inside a Generative Adversarial Network (GAN) for rapid generation of prior samples. To create a computationally efficient method we use a Mixture Density Network (MDN) to perform the inversions, and use Markov-chain Monte Carlo inversion to validate the results. We successfully inject geological prior information using the GAN, and the rapid MDN posterior estimates closely approximates the benchmark McMC posterior estimates.

Furthermore, we are able to analyse the effects of using inappropriate prior information for a given set of travel times (travel times from a braided river system inverted using prior information from marine parasequences, and vice versa). We find that the posterior estimates with inappropriate prior information are worse compared to appropriate prior information, as expected. However, the posterior estimates still have some information about the underlying true parameter matrix, so we train a neural network to find the posterior class probability that describes which prior conceptual model information to use for a given set of travel time data. We thus demonstrate that we can rapidly invert tomographic travel times with rich geological prior information, and that we are able to discriminate between a set of geological conceptual models to find which is most appropriate for the are under consideration.

\section{Acknowledgement}
We thank the sponsors of the Edinburgh Imaging Project (blogs.ed.ac.uk/imaging): TotalEnergies and BP for enabling this study. This work has made use of the resources provided by the Edinburgh Compute and Data Facility (ECDF) (http://www.ecdf.ed.ac.uk/). Furthermore, we would like to thank Niklas Linde for directing us to the river network GAN published by \cite{laloy2018training}. For the purpose of open access, the authors have applied a Creative Commons Attribution (CC BY) license to any Author Accepted Manuscript version arising.

\bibliographystyle{apalike}  

\bibliography{references}

\begin{thebibliography}{}

\bibitem[Aki and Lee, 1976]{aki1976determination}
Aki, K. and Lee, W. (1976).
\newblock Determination of three-dimensional velocity anomalies under a seismic
  array using first p arrival times from local earthquakes: 1. a homogeneous
  initial model.
\newblock {\em Journal of Geophysical research}, 81(23):4381--4399.

\bibitem[Arjovsky et~al., 2017]{arjovsky2017wasserstein}
Arjovsky, M., Chintala, S., and Bottou, L. (2017).
\newblock Wasserstein generative adversarial networks.
\newblock In {\em International conference on machine learning}, pages
  214--223. PMLR.

\bibitem[Arnold et~al., 2019]{arnold2019uncertainty}
Arnold, D., Demyanov, V., Rojas, T., and Christie, M. (2019).
\newblock Uncertainty quantification in reservoir prediction: part 1—model
  realism in history matching using geological prior definitions.
\newblock {\em Mathematical Geosciences}, 51(2):209--240.

\bibitem[Arnold and Curtis, 2018]{arnold2018interrogation}
Arnold, R. and Curtis, A. (2018).
\newblock Interrogation theory.
\newblock {\em Geophysical Journal International}, 214(3):1830--1846.

\bibitem[Bishop and Nasrabadi, 2006]{bishop2006pattern}
Bishop, C.~M. and Nasrabadi, N.~M. (2006).
\newblock {\em Pattern recognition and machine learning}, volume~4.
\newblock Springer.

\bibitem[Bodin and Sambridge, 2009]{bodin2009seismic}
Bodin, T. and Sambridge, M. (2009).
\newblock Seismic tomography with the reversible jump algorithm.
\newblock {\em Geophysical Journal International}, 178(3):1411--1436.

\bibitem[Bond et~al., 2012]{bond2012makes}
Bond, C., Lunn, R., Shipton, Z., and Lunn, A. (2012).
\newblock What makes an expert effective at interpreting seismic images?
\newblock {\em Geology}, 40(1):75--78.

\bibitem[Bond et~al., 2015]{bond2015structural}
Bond, C.~E., Johnson, G., and Ellis, J. (2015).
\newblock Structural model creation: the impact of data type and creative space
  on geological reasoning and interpretation.
\newblock {\em Geological Society, London, Special Publications},
  421(1):83--97.

\bibitem[Brock et~al., 2018]{brock2018large}
Brock, A., Donahue, J., and Simonyan, K. (2018).
\newblock Large scale gan training for high fidelity natural image synthesis.
\newblock {\em arXiv preprint arXiv:1809.11096}.

\bibitem[Burgess et~al., 2001]{burgess2001numerical}
Burgess, P., Wright, V., and Emery, D. (2001).
\newblock Numerical forward modelling of peritidal carbonate parasequence
  development: implications for outcrop interpretation.
\newblock {\em Basin Research}, 13(1):1--16.

\bibitem[Burgess and Emery, 2004]{burgess2004sensitive}
Burgess, P.~M. and Emery, D.~J. (2004).
\newblock Sensitive dependence, divergence and unpredictable behaviour in a
  stratigraphic forward model of a carbonate system.
\newblock {\em Geological Society, London, Special Publications},
  239(1):77--94.

\bibitem[Curtis, 2012]{curtis2012science}
Curtis, A. (2012).
\newblock The science of subjectivity.
\newblock {\em Geology}, 40(1):95--96.

\bibitem[Curtis and Lomax, 2001]{curtis2001prior}
Curtis, A. and Lomax, A. (2001).
\newblock Prior information, sampling distributions, and the curse of
  dimensionality.
\newblock {\em Geophysics}, 66(2):372--378.

\bibitem[Curtis and Wood, 2004]{curtis2004optimal}
Curtis, A. and Wood, R. (2004).
\newblock Optimal elicitation of probabilistic information from experts.
\newblock {\em Geological Society, London, Special Publications},
  239(1):127--145.

\bibitem[Dziewonski and Woodhouse, 1987]{dziewonski1987global}
Dziewonski, A.~M. and Woodhouse, J.~H. (1987).
\newblock Global images of the earth's interior.
\newblock {\em Science}, 236(4797):37--48.

\bibitem[Earp and Curtis, 2020]{earp2020probabilistic}
Earp, S. and Curtis, A. (2020).
\newblock Probabilistic neural network-based 2d travel-time tomography.
\newblock {\em Neural Computing and Applications}, 32(22):17077--17095.

\bibitem[Earp et~al., 2020]{earp2020grane}
Earp, S., Curtis, A., Zhang, X., and Hansteen, F. (2020).
\newblock Probabilistic neural network tomography across grane field (north
  sea) from surface wave dispersion data.
\newblock {\em Geophysical Journal International}, 223(3):1741--1757.

\bibitem[Feng et~al., 2018]{feng2018reservoir}
Feng, R., Luthi, S.~M., Gisolf, D., and Angerer, E. (2018).
\newblock Reservoir lithology classification based on seismic inversion results
  by hidden markov models: Applying prior geological information.
\newblock {\em Marine and Petroleum Geology}, 93:218--229.

\bibitem[Galetti et~al., 2015]{galetti2015uncertainty}
Galetti, E., Curtis, A., Meles, G.~A., and Baptie, B. (2015).
\newblock Uncertainty loops in travel-time tomography from nonlinear wave
  physics.
\newblock {\em Physical review letters}, 114(14):148501.

\bibitem[Georgiadis et~al., 2013]{georgiadis2013texture}
Georgiadis, G., Chiuso, A., and Soatto, S. (2013).
\newblock Texture compression.
\newblock In {\em 2013 Data Compression Conference}, pages 221--230. IEEE.

\bibitem[Gonz{\'a}lez et~al., 2008]{gonzalez2008seismic}
Gonz{\'a}lez, E.~F., Mukerji, T., and Mavko, G. (2008).
\newblock Seismic inversion combining rock physics and multiple-point
  geostatistics.
\newblock {\em Geophysics}, 73(1):R11--R21.

\bibitem[Goodfellow, 2016]{goodfellow2016nips}
Goodfellow, I. (2016).
\newblock Nips 2016 tutorial: Generative adversarial networks.
\newblock {\em arXiv preprint arXiv:1701.00160}.

\bibitem[Goodfellow et~al., 2014]{goodfellow2014generative}
Goodfellow, I., Pouget-Abadie, J., Mirza, M., Xu, B., Warde-Farley, D., Ozair,
  S., Courville, A., and Bengio, Y. (2014).
\newblock Generative adversarial nets.
\newblock {\em Advances in neural information processing systems}, 27.

\bibitem[Gulrajani et~al., 2017]{gulrajani2017improved}
Gulrajani, I., Ahmed, F., Arjovsky, M., Dumoulin, V., and Courville, A. (2017).
\newblock Improved training of wasserstein gans.
\newblock {\em arXiv preprint arXiv:1704.00028}.

\bibitem[Hill et~al., 2009]{hill2009modeling}
Hill, J., Tetzlaff, D., Curtis, A., and Wood, R. (2009).
\newblock Modeling shallow marine carbonate depositional systems.
\newblock {\em Computers \& Geosciences}, 35(9):1862--1874.

\bibitem[Jetchev et~al., 2016]{jetchev2016texture}
Jetchev, N., Bergmann, U., and Vollgraf, R. (2016).
\newblock Texture synthesis with spatial generative adversarial networks.
\newblock {\em arXiv preprint arXiv:1611.08207}.

\bibitem[Kaipio and Somersalo, 2007]{kaipio2007statistical}
Kaipio, J. and Somersalo, E. (2007).
\newblock Statistical inverse problems: discretization, model reduction and
  inverse crimes.
\newblock {\em Journal of computational and applied mathematics},
  198(2):493--504.

\bibitem[Kang and Park, 2020]{kang2020ContraGAN}
Kang, M. and Park, J. (2020).
\newblock {ContraGAN: Contrastive Learning for Conditional Image Generation}.
\newblock In {\em Conference on Neural Information Processing Systems
  (NeurIPS)}.

\bibitem[Kang et~al., 2021]{kang2021ReACGAN}
Kang, M., Shim, W., Cho, M., and Park, J. (2021).
\newblock {Rebooting ACGAN: Auxiliary Classifier GANs with Stable Training}.
\newblock In {\em Conference on Neural Information Processing Systems
  (NeurIPS)}.

\bibitem[Kass and Wasserman, 1996]{kass1996selection}
Kass, R.~E. and Wasserman, L. (1996).
\newblock The selection of prior distributions by formal rules.
\newblock {\em Journal of the American statistical Association},
  91(435):1343--1370.

\bibitem[Laloy et~al., 2018]{laloy2018training}
Laloy, E., H{\'e}rault, R., Jacques, D., and Linde, N. (2018).
\newblock Training-image based geostatistical inversion using a spatial
  generative adversarial neural network.
\newblock {\em Water Resources Research}, 54(1):381--406.

\bibitem[Lee et~al., 1995]{lee1995time}
Lee, D.~S., Stevenson, V.~M., Johnston, P.~F., and Mullen, C. (1995).
\newblock Time-lapse crosswell seismic tomography to characterize flow
  structure in the reservoir during the thermal stimulation.
\newblock {\em Geophysics}, 60(3):660--666.

\bibitem[Lochb{\"u}hler et~al., 2015]{lochbuhler2015summary}
Lochb{\"u}hler, T., Vrugt, J.~A., Sadegh, M., and Linde, N. (2015).
\newblock Summary statistics from training images as prior information in
  probabilistic inversion.
\newblock {\em Geophysical Journal International}, 201(1):157--171.

\bibitem[Moja et~al., 2019]{moja2019bayesian}
Moja, S.~S., Asfaw, Z.~G., and Omre, H. (2019).
\newblock Bayesian inversion in hidden markov models with varying marginal
  proportions.
\newblock {\em Mathematical Geosciences}, 51(4):463--484.

\bibitem[Mosegaard and Tarantola, 1995]{mosegaard1995monte}
Mosegaard, K. and Tarantola, A. (1995).
\newblock Monte carlo sampling of solutions to inverse problems.
\newblock {\em Journal of Geophysical Research: Solid Earth},
  100(B7):12431--12447.

\bibitem[Mosser et~al., 2020]{mosser2020stochastic}
Mosser, L., Dubrule, O., and Blunt, M.~J. (2020).
\newblock Stochastic seismic waveform inversion using generative adversarial
  networks as a geological prior.
\newblock {\em Mathematical Geosciences}, 52(1):53--79.

\bibitem[Otoo and Hodgetts, 2021]{otoo2021porosity}
Otoo, D. and Hodgetts, D. (2021).
\newblock Porosity and permeability prediction through forward stratigraphic
  simulations using gpm™ and petrel™: application in shallow marine
  depositional settings.
\newblock {\em Geoscientific Model Development}, 14(4):2075--2095.

\bibitem[Paola, 2000]{paola2000quantitative}
Paola, C. (2000).
\newblock Quantitative models of sedimentary basin filling.
\newblock {\em Sedimentology}, 47:121--178.

\bibitem[Podvin and Lecomte, 1991]{podvin1991}
Podvin, P. and Lecomte, I. (1991).
\newblock Finite difference computation of traveltimes in very contrasted
  velocity models: a massively parallel approach and its associated tools.
\newblock {\em Geophysical Journal International}, 105(1):271--284.

\bibitem[Polson and Curtis, 2010]{polson2010dynamics}
Polson, D. and Curtis, A. (2010).
\newblock Dynamics of uncertainty in geological interpretation.
\newblock {\em Journal of the Geological Society}, 167(1):5--10.

\bibitem[Rawlinson and Sambridge, 2004]{rawlinson2004multiple}
Rawlinson, N. and Sambridge, M. (2004).
\newblock Multiple reflection and transmission phases in complex layered media
  using a multistage fast marching method.
\newblock {\em Geophysics}, 69(5):1338--1350.

\bibitem[Richard and Lippmann, 1991]{richard1991neural}
Richard, M.~D. and Lippmann, R.~P. (1991).
\newblock Neural network classifiers estimate bayesian a posteriori
  probabilities.
\newblock {\em Neural computation}, 3(4):461--483.

\bibitem[Salimans et~al., 2016]{salimans2016improved}
Salimans, T., Goodfellow, I., Zaremba, W., Cheung, V., Radford, A., and Chen,
  X. (2016).
\newblock Improved techniques for training gans.
\newblock {\em Advances in neural information processing systems},
  29:2234--2242.

\bibitem[Song et~al., 2021a]{song2021bridging}
Song, S., Mukerji, T., and Hou, J. (2021a).
\newblock Bridging the gap between geophysics and geology with generative
  adversarial networks.
\newblock {\em IEEE Transactions on Geoscience and Remote Sensing}, 60:1--11.

\bibitem[Song et~al., 2021b]{song2021gansim}
Song, S., Mukerji, T., and Hou, J. (2021b).
\newblock Gansim: Conditional facies simulation using an improved progressive
  growing of generative adversarial networks (gans).
\newblock {\em Mathematical Geosciences}, 53(7):1413--1444.

\bibitem[Tarantola, 2005]{tarantola2005inverse}
Tarantola, A. (2005).
\newblock {\em Inverse problem theory and methods for model parameter
  estimation}, volume~89.
\newblock siam.

\bibitem[Tarantola and Valette, 1982]{tarantola1982}
Tarantola, A. and Valette, B. (1982).
\newblock Inverse problems=quest for information.
\newblock {\em J. geophys}, 50(3):150--170.

\bibitem[Tetzlaff, 2022]{tetzlaffsedsimple}
Tetzlaff, D. (2022).
\newblock \url{https://wsoftc.com}.

\bibitem[Tetzlaff and Harbaugh, 1989]{tetzlaff1989simulating}
Tetzlaff, D.~M. and Harbaugh, J.~W. (1989).
\newblock {\em Simulating clastic sedimentation}.
\newblock New York, NY; Van Nostrand Reinhold Co. Inc.

\bibitem[Tsekhmistrenko et~al., 2021]{tsekhmistrenko2021tree}
Tsekhmistrenko, M., Sigloch, K., Hosseini, K., and Barruol, G. (2021).
\newblock A tree of indo-african mantle plumes imaged by seismic tomography.
\newblock {\em Nature Geoscience}, 14(8):612--619.

\bibitem[Zhang and Curtis, 2020]{zhang2020seismic}
Zhang, X. and Curtis, A. (2020).
\newblock Seismic tomography using variational inference methods.
\newblock {\em Journal of Geophysical Research: Solid Earth},
  125(4):e2019JB018589.

\bibitem[Zhang and Curtis, 2022]{zhang2022interrogating}
Zhang, X. and Curtis, A. (2022).
\newblock Interrogating probabilistic inversion results for subsurface
  structural information.
\newblock {\em Geophysical Journal International}, 229(2):750--757.

\bibitem[Zhao et~al., 2022]{zhao2022interrogating}
Zhao, X., Curtis, A., and Zhang, X. (2022).
\newblock Interrogating subsurface structures using probabilistic tomography:
  an example assessing the volume of irish sea basins.
\newblock {\em Journal of Geophysical Research: Solid Earth},
  127(4):e2022JB024098.

\bibitem[Zhao et~al., 2020]{zhao2020improved}
Zhao, Z., Singh, S., Lee, H., Zhang, Z., Odena, A., and Zhang, H. (2020).
\newblock Improved consistency regularization for gans.
\newblock {\em arXiv preprint arXiv:2002.04724}.

\end{thebibliography}

\end{document}